\newcommand{\Msun}{\mathrm{M}_{\odot}}

\documentclass[sort&compress
    ,final            
    ,numberedheadings 
  ]
  {aipproc}

\usepackage{amssymb,amsmath,url}
\layoutstyle{6x9}

\usepackage[hang,flushmargin]{footmisc}

\begin{document}

\title[Simulating Reionization and the Radiation Backgrounds from the EOR]{Simulating Cosmic Reionization and the Radiation Backgrounds
from the Epoch of Reionization}


\author{Shapiro, Paul R.\footnotemark$^{~,}$\footnotetext{~~Invited Speaker}}{
address={Department of Astronomy and Texas Cosmology Center, University of Texas, Austin, TX 78712-1083, U.S.A.}
}

\author{Iliev, Ilian, T.}{
address={Astronomy Centre, Department of Physics and Astronomy, Pevensey II Building, University of Sussex, Falmer, Brighton BN1 9QH}
}

\author{Mellema, G.}{
address={Department of Astronomy \& Oskar Klein Centre, AlbaNova, Stockholm University, SE-106 91 Stockholm, Sweden}
}

\author{Ahn, Kyungjin}{
address={Department of Earth Sciences, Chosun University, Gwangju 501-759, Republic Korea}
}

\author{Mao, Yi}{
address={Department of Astronomy and Texas Cosmology Center, University of Texas, Austin, TX 78712-1083, U.S.A.}
}

\author{Friedrich, Martina}{
address={Department of Astronomy \& Oskar Klein Centre, AlbaNova, Stockholm University, SE-106 91 Stockholm, Sweden}
} 

\author{Datta, Kanan}{
address={Department of Astronomy \& Oskar Klein Centre, AlbaNova, Stockholm University, SE-106 91 Stockholm, Sweden}
}

\author{Park, Hyunbae}{
address={Department of Astronomy and Texas Cosmology Center, University of Texas, Austin, TX 78712-1083, U.S.A.}
}

\author{Komatsu, Eiichiro}{
address={Department of Astronomy, University of Texas, Austin, TX 78712-1083, U.S.A.}
}

\author{Fernandez, Elizabeth}{
address={Univ Paris-Sud, Institut d'Astrophysique Spatiale, UMR8617, 91405 Orsay Cedex, France}
}

\author{Koda, Jun}{
address={Centre for Astrophysics and Supercomputing, Swinburne University of Technology, Hawthorn, Victoria 3122, Australia}
}

\author{Bovill, Mia}{
address={Department of Astronomy and Texas Cosmology Center, University of Texas, Austin, TX 78712-1083, U.S.A.}
}

\author{Pen, Ue-Li}{
address={Canadian Institute for Theoretical Astrophysics, University
of Toronto, 60 St. George Street, Toronto, ON M5S 3H8, Canada}
}

\begin{abstract}
Large-scale reionization simulations are described which combine the results of cosmological N-body simulations that model the evolving density and velocity
fields and identify the galactic halo sources, with ray-tracing radiative transfer calculations which model the nonequilibrium ionization of the intergalactic medium.  
These simulations have been used to predict some of the signature effects of reionization on cosmic radiation backgrounds, including the CMB, near-IR, and redshifted 21cm backgrounds.  We summarize some of our recent progress in this work, and address the question of whether observations of such signature effects can be used to distinguish the relative contributions of galaxies of different masses to reionization.
\end{abstract}

\maketitle

\section{Introduction}

The talk on which this paper is based summarized some of our recent progress towards simulating cosmic reionization and the radiation backgrounds from this epoch \footnote{\url{http://tpweb2.phys.konan-u.ac.jp/~FirstStar4/presentation_files/PShapiro.pdf}}.  Here we are constrained by length limitations to be even more selective, so we will not presume to review this large and active field, but rather will focus on just a few of our most recent results mentioned in the talk. 

When the first stars and galaxies formed, they released ionizing radiation into the surrounding intergalactic medium (IGM), creating a patchwork quilt of H II regions
that grew in size and number and merged over time, until their overlap eventually   
ionized the entire IGM.  The spectra of distant quasars and galaxies seen through this IGM suggest this overlap occurred by redshift $z_{ov} \gtrsim 6$, while fluctuations in the polarization of the CMB on large angular scales imply that reionization must have begun at significantly higher redshift, to provide enough intergalactic, electron-scattering optical depth, $\tau_{es}$, to explain this.  Observing cosmic reionization, then, would seem to be a good way to learn about first star and galaxy formation. While direct observation of the sources of reionization remains a challenge, their impact on cosmic background radiation during the epoch of reionization (EOR) offers another probe.    

Predictions require the modelling of the inhomogeneity of reionization on very large scales (>~ 100/h cMpc), starting from realistic cosmological initial conditions, in order to be statistically meaningful.  Ideally, such models would simultaneously resolve the formation and internal structure of individual galaxies and, inside each, the births and deaths of all the stars, including exchange of mass, energy and radiation between the IGM and the galaxies and the back-reaction caused by reionization, itself.  The dynamic range required for this "ideal" is currently very far from achievable, however. 

Fortunately, there is some separation of scales which is possible as a first approximation, in which the location sites of reionization sources are identified with the dark-matter-dominated galactic halos that host galaxies, their efficiency as sources is parameterized in some phenomenological way, and their radiation is transferred through the evolving IGM, whose changing ionization state also changes its opacity.  This approximation is aided by the fact that reionization proceeds by propagating ionization fronts (I-fronts) that travel highly supersonically through the IGM, the I-front regime known as "weak, R-type," so the H II region boundaries race ahead of the hydrodynamical back-reaction of the gas in response to its pressure increase by photoheating \cite{1987ApJ...321L.107S}.  While this back-reaction can affect small-scale structure, it does not directly affect the large-scale pattern of reionization patchiness.  On the other hand, it does affect the efficiency of small-mass galaxies as sources, since the pressure increase of the photoheated IGM opposes baryonic gravitational collapse on small-scales, sometimes referred to as "Jeans-mass filtering"  \cite{1994ApJ...427...25S,2000ApJ...542..535G}, so small-mass halos inside IGM H II regions may have significantly reduced star formation rates.  By changing the source efficiencies in this way, such feedback can alter the large-scale structure of reionization, too.   The mass upper-limit of halo sources suppressed this way is uncertain, but
estimates range at least to $10^9\Msun$.  The smallest galactic halos that can form stars, minihalos with mass between $10^5$ and $10^8$ $\Msun$ (with virial temperatures $T< 10^4$K), are especially vulnerable to such radiative suppression inside H II regions, since they can also lose the interstellar gas they already have, if overtaken by an I-front as it crosses the IGM, which expels their gas by photoevaporation \cite{shapiro/iliev/raga:2004}.  Minihalo star formation can also be suppressed by the rising UV background below 13.6 eV emitted by the same reionization sources, of H$_2$-dissociating radiation in the Lyman-Werner bands (e.g.\cite{2000ApJ...534...11H}), about which we will say more below, which limits their contribution to reionization even in neutral patches of the IGM.  Minihalos depend upon H$_2$ molecules to cool their gas below their virial temperatures to enable star formation, since atomic cooling in primordial composition gas is ineffective below $10^4$K.  

In what follows, therefore, we will group the sources of reionization according to their halo mass, in three broad bins: MiniHalos (MH), Low-Mass Atomic-Cooling Halos (LMACH)
with $10^8 < M < 10^9$ $\Msun$, and High-Mass Atomic-Cooling Halos (HMACH)
above $10^9 \Msun$. MH and LMACH sources are both suppressed inside H II regions, MHs are suppressed in neutral zones, too, if the mean intensity of LW background radiation there exceeds some threshold value, $J_{\mathrm{LW}} > J_{\mathrm{LW,th}}$, and HMACHs are immune to suppression.    We will say more about the relative contributions of these
three source types to reionization below, but first we describe the simulations we
will apply to address that question.

The numerical methods we have developed for this purpose are described in Section 2. If this were a review of the field of reionization simulation, I would also mention
the large body of work by other authors, too, that has gone into developing cosmological radiative transfer methods and applying them to this problem, on scales large and small, apart from our own, including \cite{2006PASJ...58..445S,2011MNRAS.414.3458W,2008MNRAS.389..651P,2011MNRAS.412..935P,2007A&amp;A...474..365S,2012MNRAS.419.2855O,2011MNRAS.411.1678C,2012MNRAS.422.3067P,2009MNRAS.396.1383P,2009MNRAS.393.1090F,2008MNRAS.386.1931A,2008MNRAS.388.1501C,2011ASL.....4..228T,2010ApJ...724..244A,2008MNRAS.387..295A,2009A&amp;A...495..389B,2007ApJ...671....1T,2008ApJ...681..756S,2007MNRAS.377.1043M,2006ApJ...639..621A,2005ApJ...633..552K,2009MNRAS.393..171M,2003MNRAS.344..607S,2002ApJ...572..695R,2003MNRAS.345..379M,2001NewA....6..437G,1999ApJ...523...66A,2012MNRAS.423..558C,2003MNRAS.343.1101C,2000ApJ...535..530G}.  To test these methods, we organized the \emph{Cosmological Radiative Transfer Comparison Project} several years ago, with workshops leading to published test problems and comparisons which have encouraged this rapid growth \cite{iliev/etal:2006a,iliev/etal:2009}.  

\section{Reionization Simulations: N-body + Radiative Transfer}

\subsection{N-body Simulations}

For our second generation of reionization simulations, the density and velocity fields of the IGM and the source halos are given by large-scale cosmological N-body simulations by the new N-body code CUBEP$^3$M, a P$^3$M method, massively paralleled (MPI + Open MP)
(\cite{iliev/etal:2006b,2005NewA...10..393M}, J. Harnois-Deraps et al. 2012, in preparation).  We mention three new simulations here with large particle number $N$, on a grid of N$_{\mathrm{cells}}$, in comoving periodic box of size $L_{\mathrm{box}}$, particle mass $m_p$, and minimum resolved halo mass $M_{\mathrm{min}}$, given by ($N$, $N_{\mathrm{cell}}$, $L_{\mathrm{box}}$, $m_p$, $M_{\mathrm{min}}$) = ($3073^3$, $6,144^3$, $163$ Mpc, $5\times 10^6\Msun$, $10^8\Msun$), ($5488^3$, $10,976^3$, $29$ Mpc, $5 \times 10^3\Msun$, $10^5\Msun$), and ($5488^3$, $10,976^3$, $607$ Mpc, $5 \times 10^7\Msun$, $10^9\Msun$), respectively.  Halos are identified "on-the-fly" by a spherical overdensity halo finder.  Of these, the $163$ ($= 114/h$) Mpc simulation with
29 billion particles resolves all LMACHS and HMACHs (i.e. $ \geq 10^8\Msun$), of which there are more than $10^7$ by $z\sim8$.  The $29$ ($= 20/h$) Mpc simulation resolves MHs 
($\geq 10^5\Msun$), the first of which form at $z = 43$, with more than $10^8$ by $z=8$, with length resolution as small as $182$ pc.  The 607 ($=425/h$) Mpc box resolves all the HMACHS ($\geq10^9\Msun$), the first forming at $z = 26$, with $4 \times 10^7$ halos by $z=8$, in a volume comparable to that of the LOFAR EOR 21cm background survey.

\subsection{Radiative Transfer}

Radiative transfer (RT) simulations evolve the radiation field and nonequilibrium ionization state of the gas, in the density field provided by the N-body results above, smoothed to a coarser grid of  $\sim 256^3$ to $512^3$ cells, for different resolution runs, by ray-tracing the ionizing radiation from every galaxy halo in the box.  A new, fast and efficient method was developed, the C$^2$-Ray code (Conservative, Causal Ray-Tracing),
which uses short characteristics and has the advantage that it accurately tracks I-fronts even using relatively large cell sizes (i.e. very optically thick when neutral) and time steps (i.e. much larger than the light-crossing and I-front crossing times per cell).   While our first generation of reionization simulations used this method coded in Open MP for shared memory parallel computers \cite{iliev/etal:2006b,iliev/etal:2008,iliev/etal:2007}, the second generation simulations highlighted here used a new version of C$^2$-Ray, recoded for massively parallel computers with distributed memory (MPI + Open MP) \cite{iliev/etal:2008b,shapiro/etal:2008,iliev/etal:2012}.  The original C$^2$-Ray only included hydrogen, but we recently generalized it to take account of helium and hard-spectrum sources (including X-rays), as well \cite{friedrich/etal:2012}, and applied this to study the effect of adding quasar sources to a simulation which is otherwise reionized by starlight \cite{datta/etal:2012b}.  To track the inhomogeneous, H$_2$-dissociating, LW background (11.2 - 13.6 eV) from reionization sources, important for its feedback effect on MHs, we developed another algorithm which sums the contributions at each cell from all sources along the past light cone of that cell in every direction, taking account of cosmological redshifting, geometric dilution, and attenuation by the intervening H Lyman series opacity of the IGM \cite{ahn/etal:2009,ahn/etal:2012}.  

Every galaxy in the simulation emits ionizing radiation (unless suppressed) with efficiency parameterized as follows.  Within each halo mass group, we assume a constant M/L for simplicity, with $f_{\gamma}$ ionizing photons released by each galaxy per halo baryon, where $f_{\gamma} = f_{*} f_{\mathrm{esc}} N_\mathrm{i}$,  and $f_*$ is the star-forming fraction of halo baryons, $f_{\mathrm{esc}}$ is the ionizing photon escape fraction, and $N_i$ is the number of ionizing photons emitted per stellar baryon over the stellar lifetime.  For example, if $N_i = 50,000$ (top-heavy IMF), $f_* = 0.2$, and $f_{\mathrm{esc}} = 0.2$, then $f_\gamma = 2000$, while if $N_i = 4,000$ (Salpeter IMF), $f_* = 0.1$, and $f_{\mathrm{esc}} = 0.1$, then $f_\gamma = 40$.  This yields a source luminosity $dN_{\gamma}/dt = f_{\gamma} M_{\mathrm{bary}}/(\mu  m_H  \Delta t_*$), for source lifetime (e.g. duration of burst) 
$\Delta t_*$ (e.g. $2 x 10^7$ years), $M_{\mathrm{bary}} =$ halo baryonic mass $= M_{\mathrm{halo}} (\Omega_{\mathrm{bary}}/ \Omega_{\mathrm{m}})$, leading to a star formation rate $\mathrm{SFR} = (f_{\gamma}/ \Delta t_*) ( M_{\mathrm{bary}}/ f_{\mathrm{esc}} N_i) \approx (1.7 \Msun/{\mathrm{yr}}) (f_{\gamma}/40)(0.1/f_{\mathrm{esc}})(4000/N_i) (10 \mathrm{Myr}/ \Delta t_*) (M_{\mathrm{halo}}/10^9 \Msun)$  [e.g. if $f_{\gamma} = 40$, $f_{\mathrm{esc}} = 0.1$, $f_* = 0.1$, and $\Delta  t_* = 2 x 10^7$, then $\mathrm{SFR} \sim (0.8 \Msun /\mathrm{yr}) (M_{\mathrm{halo}}/10^9 \Msun)$]. 
Since photons are released over the time interval $\Delta t_*$, we define a second efficiency parameter which reflects this dependence, too: $g_{\gamma} = f_{\gamma}/(\Delta t_*/10 \mathrm{Myr})$.

When MH sources are included, we assume one massive Pop III star forms per halo
(or an equivalent group of stars with the same effective $N_i$-value) before disrupting its own ISM and preventing further star formation, at least temporarily, in that MH, and
$f_{\mathrm{esc}} = 1$.  For such stars, we take $\Delta t_* = 1.92$ Myr and $N_i = 55,000$, which implies an efficiency for halo mass $M_{\mathrm{halo}}$ and stellar mass $M_{*,III}$ given by 
$f_{\gamma,\mathrm{MH}}$($M_{\mathrm{halo}}$, $M_{*,III}$) $= 338 (M_{*,III}/100 \Msun)/(M_{\mathrm{halo}}/10^5 \Msun)$, where the average value of $f_{\gamma,\mathrm{MH}}$ integrated over the halo mass function is comparable to the value at the low mass end of star-forming MHs.  

To account for the photoionization suppression of MHs and LMACHs, we set their $f_{\gamma}$ to zero once the halos are inside H II regions.  MHs are assumed suppressed by H$_2$ dissociation even in neutral zones, if they are located where $J_{\mathrm{LW }}> J_{\mathrm{LW,th}}$, with value chosen in the range $[0.01 - 0.1 ] \times 10^{-21} \mathrm{erg} \mathrm{s}^{-1} \mathrm{cm}^{-2} \mathrm{sr}^{-1}$, found by other work as the threshold for suppressing MH star formation.

In what follows, the reionization simulation cases are labelled to indicate the value of these parameters for different mass groups.  The full notation that appears on some plots reads ($L_{\mathrm{box}}$)\_g($g_{\gamma, \mathrm{HMACH}}$)\_($g_{\gamma, LMACH}$)S\_M($M_{*,III}$)\_J($J_{\mathrm{LW,th}}/10^{-21}$), where ``S" refers to LMACH suppression (e.g. 163Mpc\_g8.7\_130S\_M300\_J0.05).

\section{Results}

\subsection{Self-Regulated Reionization}

\cite{iliev/etal:2007} demonstrated that, when LMACHs are included but suppressed inside H II regions, reionization is "self-regulated."   The (more abundant) LMACHs start reionization earlier than the (less abundant) HMACHs, but the LMACH contribution saturates when only a fraction of the volume is ionized (i.e. long before $z_{\mathrm{ov}}$), so reionization is finished by the HMACHs, which ultimately dominate reionization.  This helped to explain how reionization might end late but extend in time enough to boost $\tau_{\mathrm{es}}$.  The N-body simulations for this had to resolve all halos with $M \geq 10^8 \Msun$, and the C$^2$-Ray simulations had to transfer the radiation from all of them, which limited the box size in practice to 50 ($= 35/h$) Mpc.  To do this for a box large enough to make statistically meaningful predictions of observables required us to develop the second generation of codes described in Section 2.  Some of the results of these self-regulated reionization simulations for box size 163 Mpc and RT grid with $256^3$ cells are described in \cite{iliev/etal:2012,shapiro/etal:2008,jensen/etal:2012,iliev/etal:2008,mao/etal:2012,datta/etal:2012,2011MNRAS.413.1353F,friedrich/etal:2012}. 

To resolve all the individual MH sources, too, in so large a volume would require an N-body simulation with more than 30 trillion particles, however, well beyond current capabilities.  Instead, we assign the MH contribution to each cell of the radiative transfer grid by a subgrid model based upon smaller-box N-body simulations like the 29 Mpc box mentioned above and a 9 Mpc box, both of which fully resolve the MH halo mass range, from which we derive the dependence of the MH mass function on local matter density, smoothed on the scale of the radiative transfer grid cells in our large-box reionization simulation \cite{ahn/etal:2012}.  With this, we have simulated reionization with MH sources, too, including their LW suppression, and found that a new kind of self-regulation occurs, in which MH sources start reionization even earlier than LMACHs, but they, too, saturate after ionizing only a fraction of the IGM.  The later rise of the ACHs ultimately stops MH star formation altogether, and reionization is once again finished (and dominated) by the HMACHs, just as without MHs \cite{ahn/etal:2012}.  With MHs, however, reionization is greatly extended, which boosts $\tau_{\mathrm{es}}$ and the large-angle polarization fluctuations of the CMB.

\subsection{Very Large-Scale Reionization Simulations: 607 Mpc box }

To simulate self-regulated reionization in an even larger volume, to test if our simulation results for the 163 Mpc box have converged yet as a function of increasing box size and enable us to predict the 21cm background for a volume comparable to that of the LOFAR EOR survey, we have simulated reionization in the 607 Mpc box described in Section 2, with $504^3$ RT cells.  Our N-body simulation directly resolves all HMACHs (i.e. $\geq 10^9 \Msun$), but we also include the LMACHs by a similar subgrid prescription to that mentioned above for the MHs, by using smaller-box simulations with enough mass resolution to resolve LMACHs directly, to derive the LMACH mass function in each radiative transfer grid cell of
the 607 Mpc box, as it varies with the cell-averaged matter density there.  Some results have already appeared in \cite{datta/etal:2012b}, in which 21cm brightness temperature fluctuation maps were made for two cases, one ionized only by starlight, while the other included a quasar source, as well, to see if a matched-filter technique applied to the 21cm observations could successfully measure the presence and size of the H II region
surrounding a quasar during the EOR.  

\begin{figure}
  \includegraphics[height=.35\textheight]{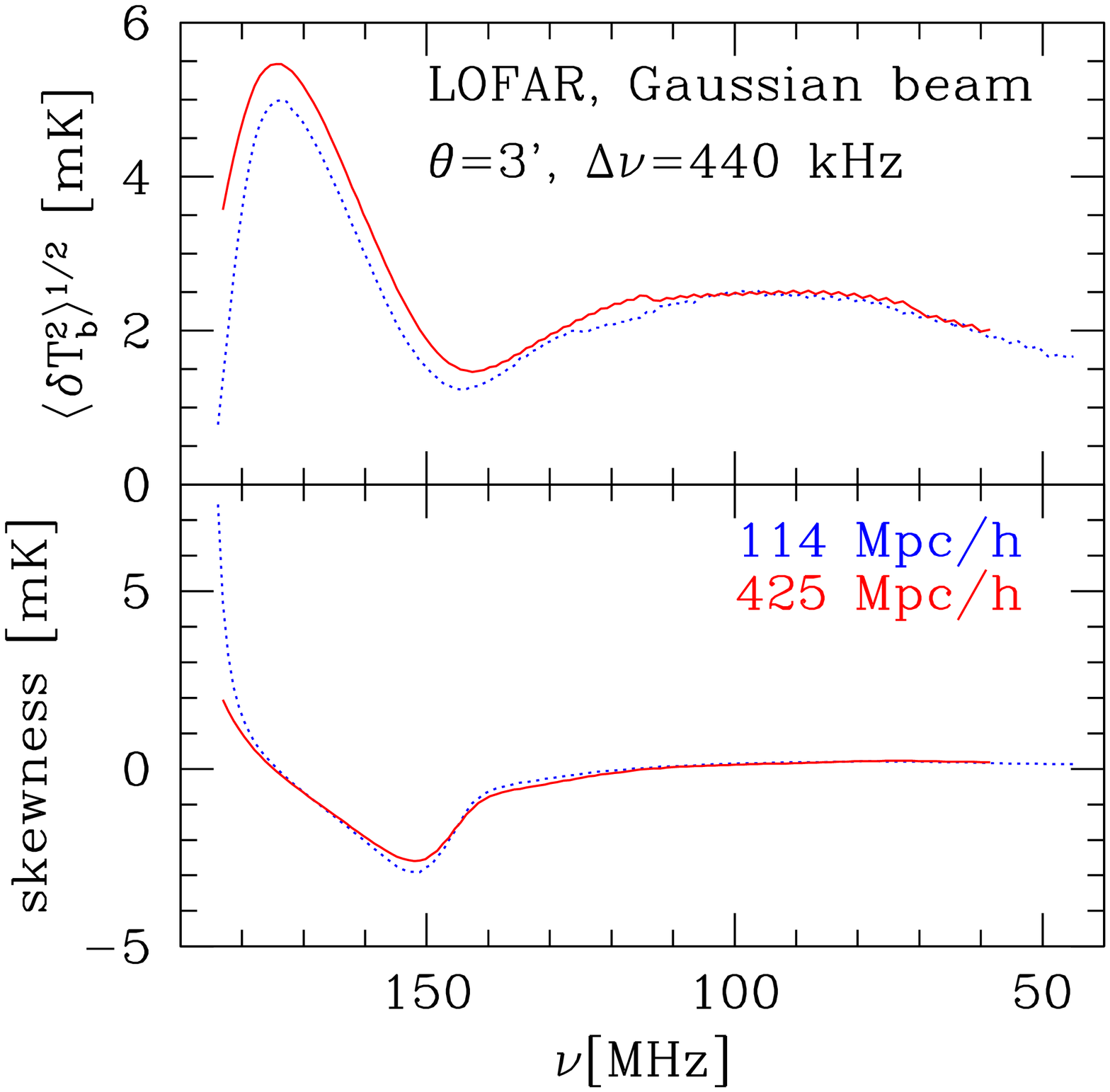}
  \includegraphics[height=.35\textheight]{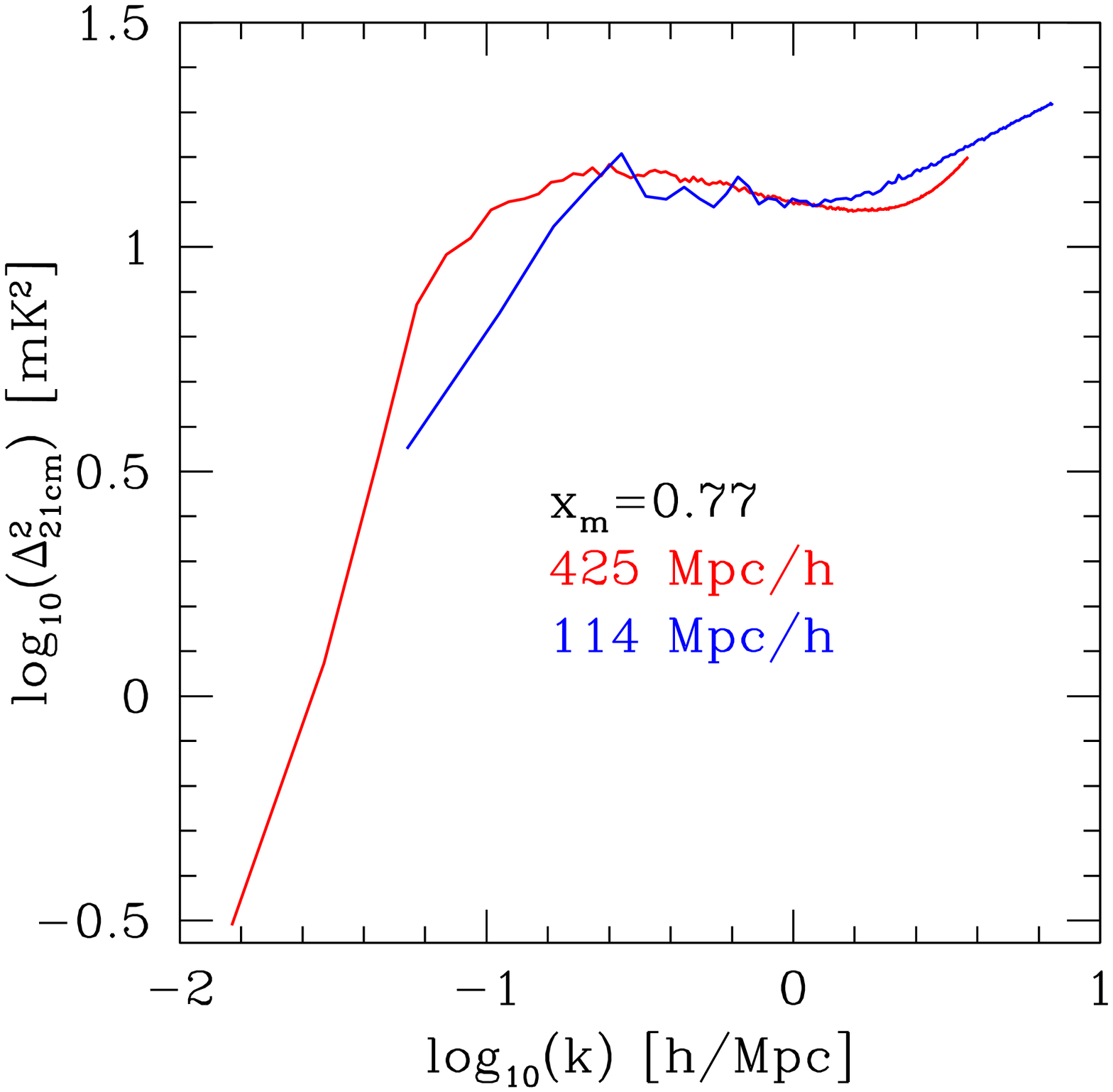}
  \caption{21cm brightness temperature fluctuations from the EOR for Case L2 (see Table 1) (HMACH+LMACH, no MHs) for 2 box sizes (163 Mpc with $256^3$ RT cells, 607 Mpc with $504^3$).  (Left panel) Rms fluctuations and skewness vs. frequency (dotted curve $=163$ Mpc).  (Right panel)  Power spectrum vs. wavenumber at $z \gtrsim 7 $, when mean ionized fraction is $x_i=0.77$ (leftmost curve $=607$ Mpc).}
\end{figure}

One comparison of interest of the 607 and 163 Mpc box simulations of self-regulated reionization is the 21cm brightness temperature predictions for each.  In the high spin temperature limit generally assumed for this EOR signal, the differential brightness temperature $\delta T_{\mathrm{b}} = T_{\mathrm{b}} - T_{\mathrm{CMB}} = (28.74 \mathrm{mK})  x_{\mathrm{HI}}  (1 + \delta) [(1+z)/10]^{1/2} \{1 + [(1+z)/H(z)](d \mathrm{v}_{\parallel}/d r_{\parallel})\}^{-1}$, where $x_{\mathrm{HI}}$ is the neutral fraction, $\delta$ is the overdensity, and $\mathrm{v}_{\parallel}$ is peculiar velocity component along the line of sight, observed at the redshifted frequency which takes peculiar velocity into account. In Figure 1 (left panel), we compare the results for rms fluctuations and skewness of the maps smoothed with beamwidth and bandwidth like those for the LOFAR survey, for the two cases, which share the same global mean ionization history, finding good agreement.  The fluctuation power spectrum is compared in Figure 1 (right panel) for the frequency which corresponds to the epoch when the global mean ionized fraction was 0.77 for both cases, showing that the larger box has more power at small wavenumber, as expected from the finite box size effect on the small box.   At higher k, however, the agreement is better.  These results will be
presented in more detail in Iliev et al. (2012, in prep).   An illustrative sky map of the 21cm brightness fluctuations from the big box simulation at observer frequency 115.827 MHz from z = 11.26 early in EOR, for a LOFAR-like beam (3 arcmin and average signal is zero) is shown in Figure 2 (left panel).

\subsection{Which Galaxies Reionized the Universe?}

\begin{table}
\caption{Reionization simulation results for the kSZ effect: global reionization history and CMB fluctuations.}
    \begin{tabular}{|l|l|l|l|l|l|l|l|}
\hline
        Case label\footnote{  L1 = 163Mpc\_g8.7\_130S = HMACH+LMACH, early reionization; \\    L2 = 163Mpc\_g1.7\_8.7S = HMACH+LMACH, late reionization; \\  L2M1 = 163Mpc\_g1.7\_8.7S\_M300\_J0.1 = HMACH+LMACH+MH; \\ L3 = 163Mpc\_g21.7\_0 = HMACH only, early reionization. \\  }         &   $z_{\mbox{ov}}$  & $z_{\mbox{99\%}}-z_{\mbox{20\%}}$ & $z_{\mbox{75\%}}-z_{\mbox{25\%}}$ & $(D_{l=3000}^{\mbox{post-reion}})\footnote{Post-reionization kSZ values are CSF model of \cite{2012ApJ...756...15S} for baryon correction.}$  & $D_{l=3000}^{\mbox{reion}}$ & $D_{l=3000}^{\mbox{total}}$ \\  \hline
        L1             & 8.3  & 2.9  & 1.8 & 1.94 $\mu\mathrm{K}^2$  & 0.83 $\mu\mathrm{K}^2$ & 2.77 $\mu\mathrm{K}^2$      \\  
        L2             & 6.7  & 1.8  & 1.2 & 1.69 $\mu\mathrm{K}^2$ & 0.66 $\mu\mathrm{K}^2$ & 2.35 $\mu\mathrm{K}^2$     \\
        L2M1       & 6.7 & 6.6 & 1.8 & 1.69 $\mu\mathrm{K}^2$ & 0.69 $\mu\mathrm{K}^2$ & 2.38  $\mu\mathrm{K}^2$     \\  
        L3             & 8.4  & 1.3  & 0.9 & 1.96 $\mu\mathrm{K}^2$  & 0.75 $\mu\mathrm{K}^2$ & 2.71 $\mu\mathrm{K}^2$    \\  \hline
    \end{tabular}
\label{table:D_l}
\end{table}

\begin{figure}
  \includegraphics[height=.35\textheight]{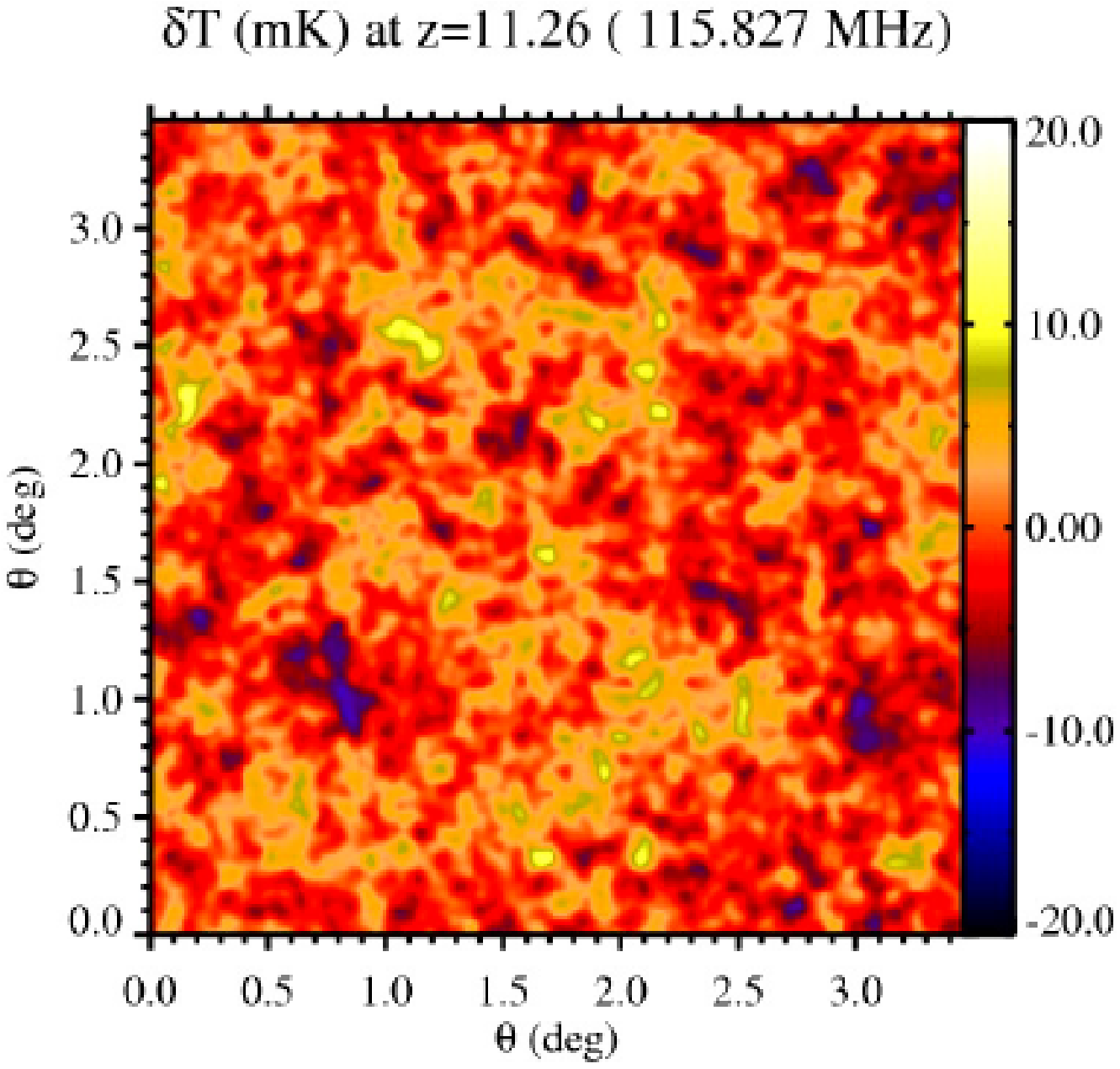}
  \includegraphics[height=.3\textheight]{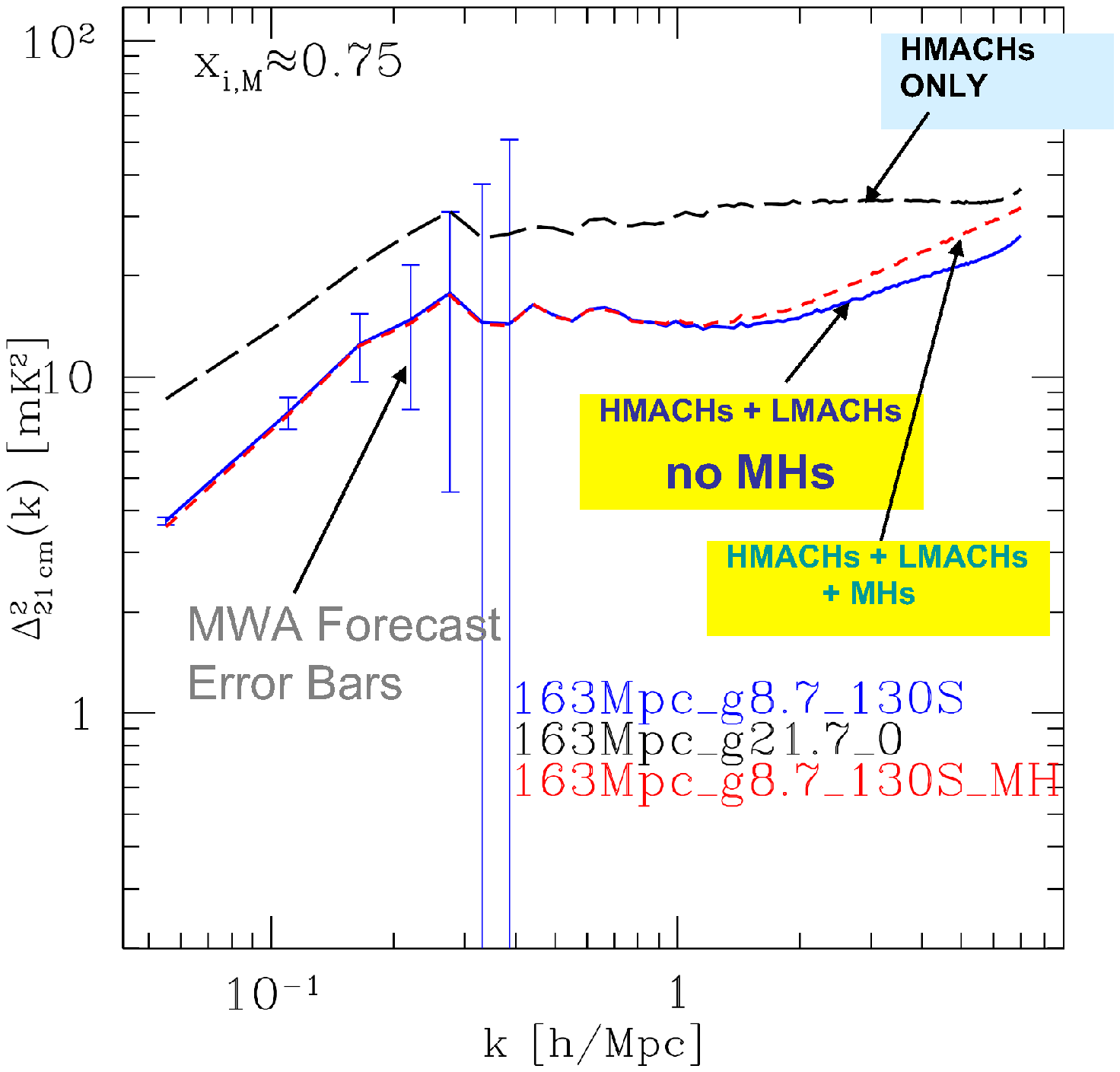}
  \caption{21cm brightness temperature fluctuations.  (Left panel) Sky map for 607 Mpc box simulation observed at 115.827 MHz (i.e. looking back to $z=11.26$, early in EOR), smoothed for LOFAR-like beam (3' and average signal is zero).  (Right panel) Comparison of power spectra at $x_{i,m}=0.75$ for 3 cases, as labelled.  }
\end{figure}

Here we use a series of 163 Mpc box simulations to ask if fluctuations in the radiation backgrounds (21cm, CMB --- polarization and kinetic Sunyaev-Zel'dovich, near-IR) from the EOR will allow us to determine which of the three types of galactic sources discussed above contribute most significantly to reionization.  To keep track of the different cases used to address this question, the reader is referred to Table 1 and the meaning of the parameter labels for different cases.  

\emph{Can 21cm fluctuations tell us which galaxies reionized the universe?}  As shown in Figure 2 (right panel), for an illustrative epoch when global ionized fraction is $x_i = 0.77$, the difference between the curves for the case with HMACHs-only and those for the cases
with LMACHS, with and without MHs, is larger than the error bars for a measurement of the power spectrum by a survey like MWA, so we might be able to distinguish the HMACH-only case, with higher power, from the cases with LMACHs, with or without MHs.  But it cannot distinguish the case of HMACH + LMACH without MHs from that with MHs.

\emph{Can CMB polarization fluctuations tell us which galaxies reionized the universe?} As shown in Figure 3 (left panel), of three cases that all finish ionizing at the same redshift $\sim 8.3$, the $\tau_{\mathrm{es}}$ for the HMACH-only case is just below the WMAP7 1-sigma range, while both cases with
LMACHs (with and without MHs) stay within that range, although the MH case has the highest $\tau_{\mathrm{es}}$.  According to Figure 3 (right panel), even the higher sensitivity of Planck after 2-years of data will distinguish the HMACHs-only case from the others, but will hardly be able to distinguish the cases with HMACHs + LMACHs, with and without MHs, from each other, for such an early end to reionization.  However, as Figure 4 (left panel) shows us,
if reionization ended as late as $z < 7$, as some other observational evidence suggests,
then the case with HMACHs + LMACHs (no MHs) makes $\tau_{\mathrm{es}}$ too small to be within the the WMAP7 1-sigma range, but HMACHs + LMACHs + MHs is within this range.  A more careful distinction requires the Planck 2-year data soon to be released, however, but
Figure 4 (right panel)  clearly shows that, even if it is within the current WMAP uncertainties, the boost to $\tau_{\mathrm{es}}$ and the polarization fluctuations caused by MHs should be readily detectable by Planck if reionization ended this late.  Planck will thereby see the signature of the first stars at high redshift, currently undetectable by any other probe  \cite{ahn/etal:2012} .

\begin{figure}
  \includegraphics[height=.3\textheight]{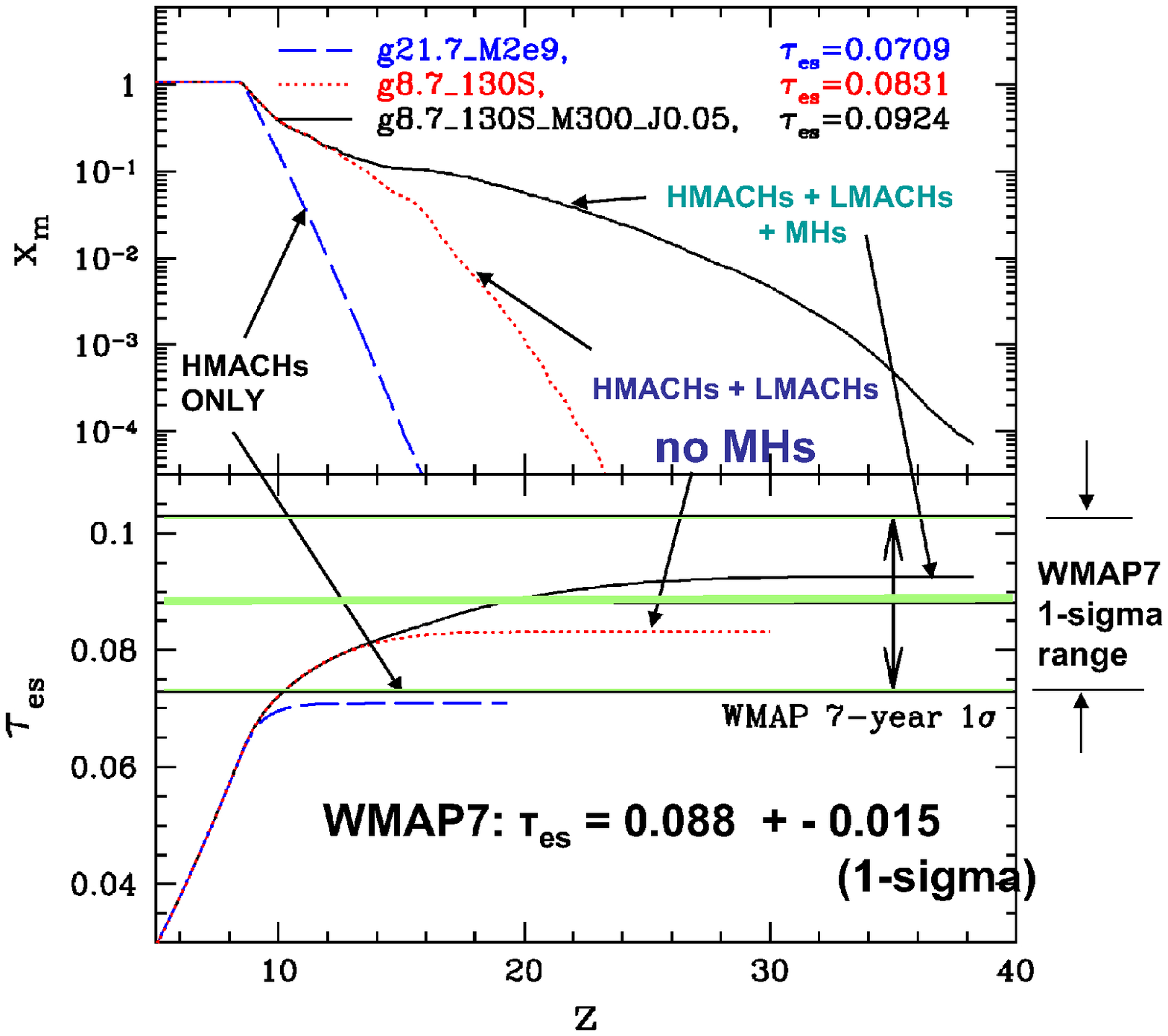}
  \includegraphics[height=.3\textheight]{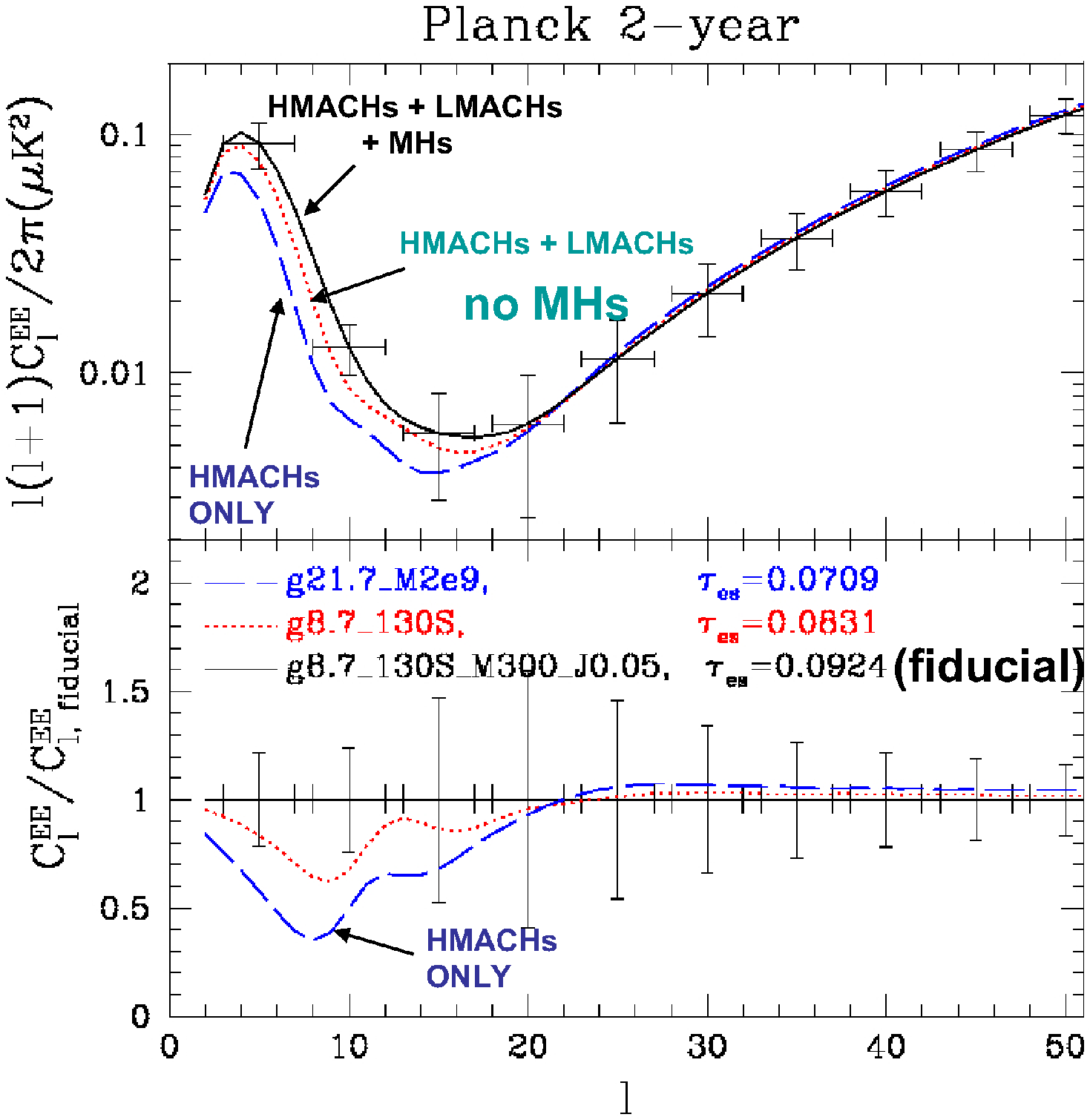}
  \caption{ (Left, top)  Global ionization histories for three cases. (Left, bottom) Electron-scattering optical depths for three cases, compared with value from CMB polarization fluctuations; (Right) Predicted CMB polarization fluctuation angular power spectra for same three cases.  Error bars are estimated Planck 2-yr, 1-$\sigma$ sensitivity including cosmic variance.  }
\end{figure}

\begin{figure}
  \includegraphics[height=.3\textheight]{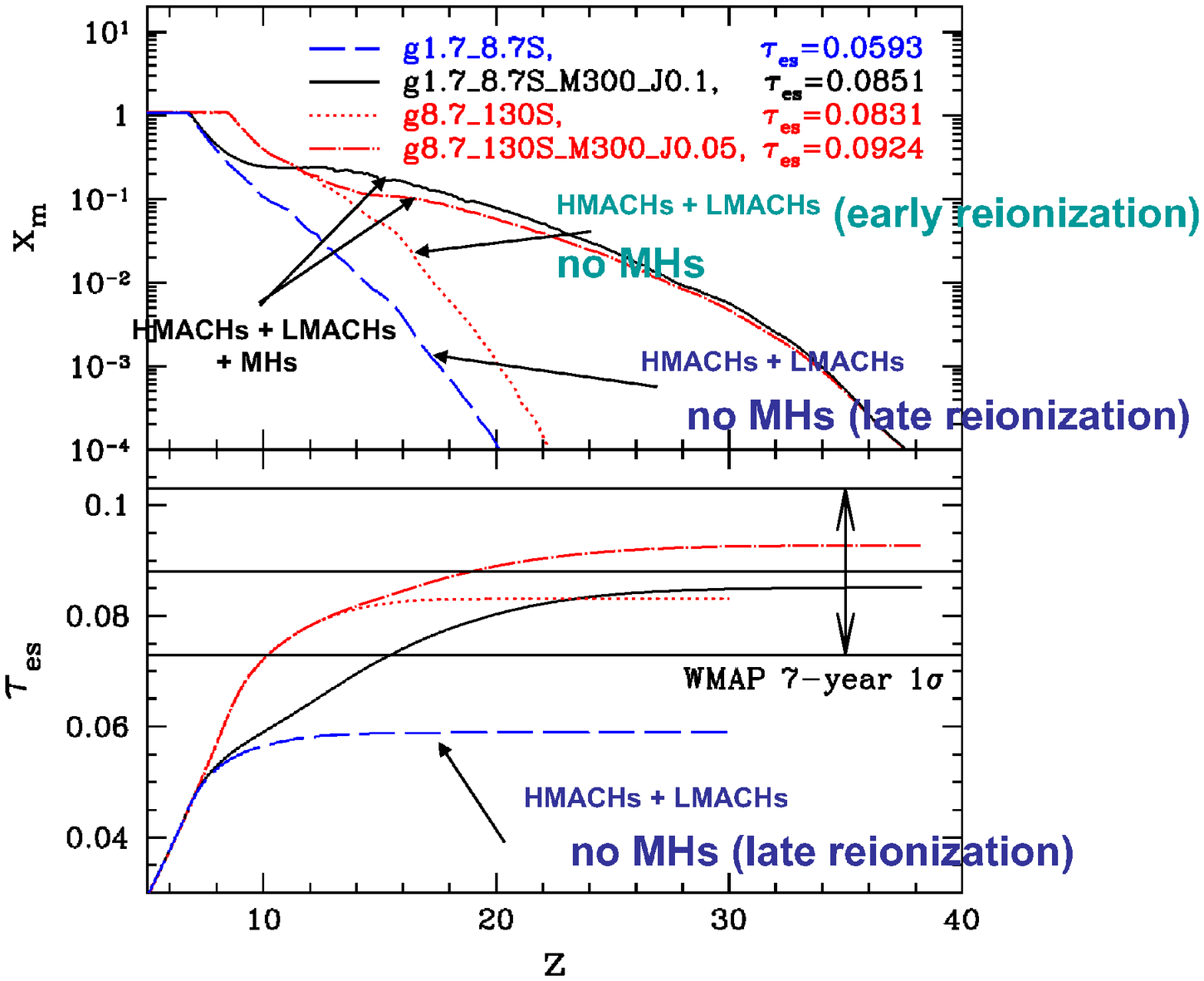}
  \includegraphics[height=.3\textheight]{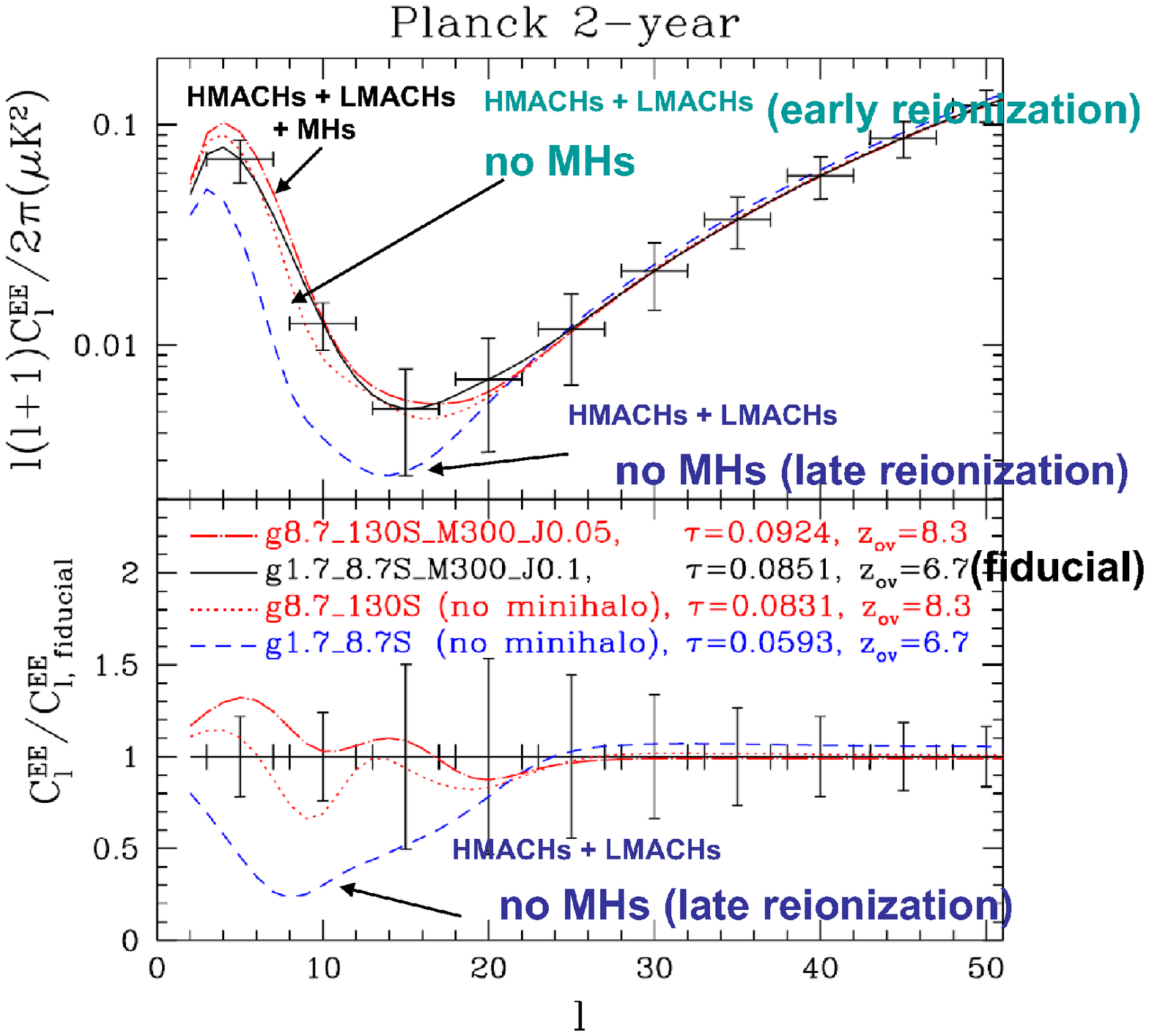}
  \caption{Same as Figure 3, except for different cases, including both early and late reionization cases.}
\end{figure}

\begin{figure}
  \includegraphics[height=.25\textheight]{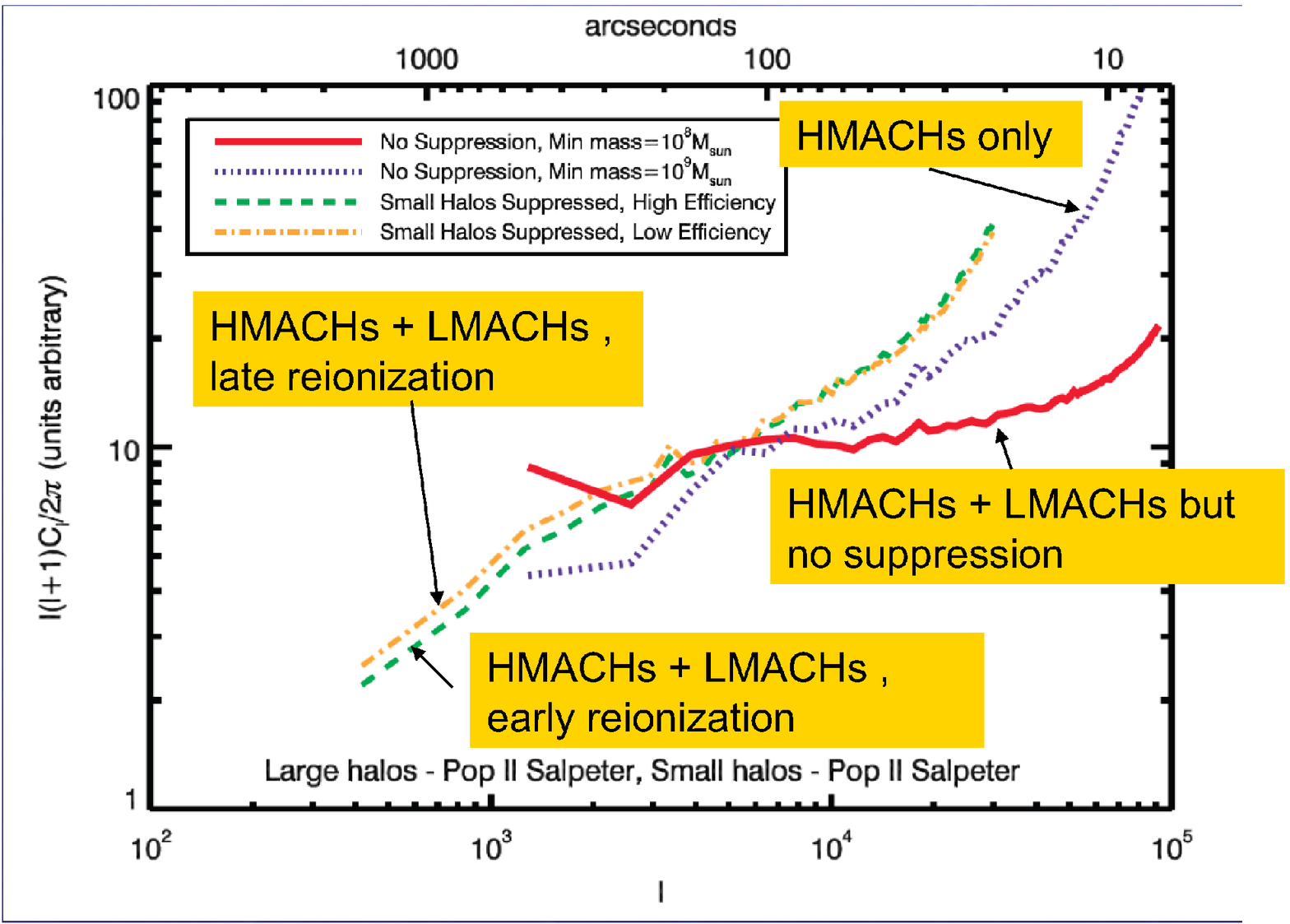}
  \includegraphics[height=.25\textheight]{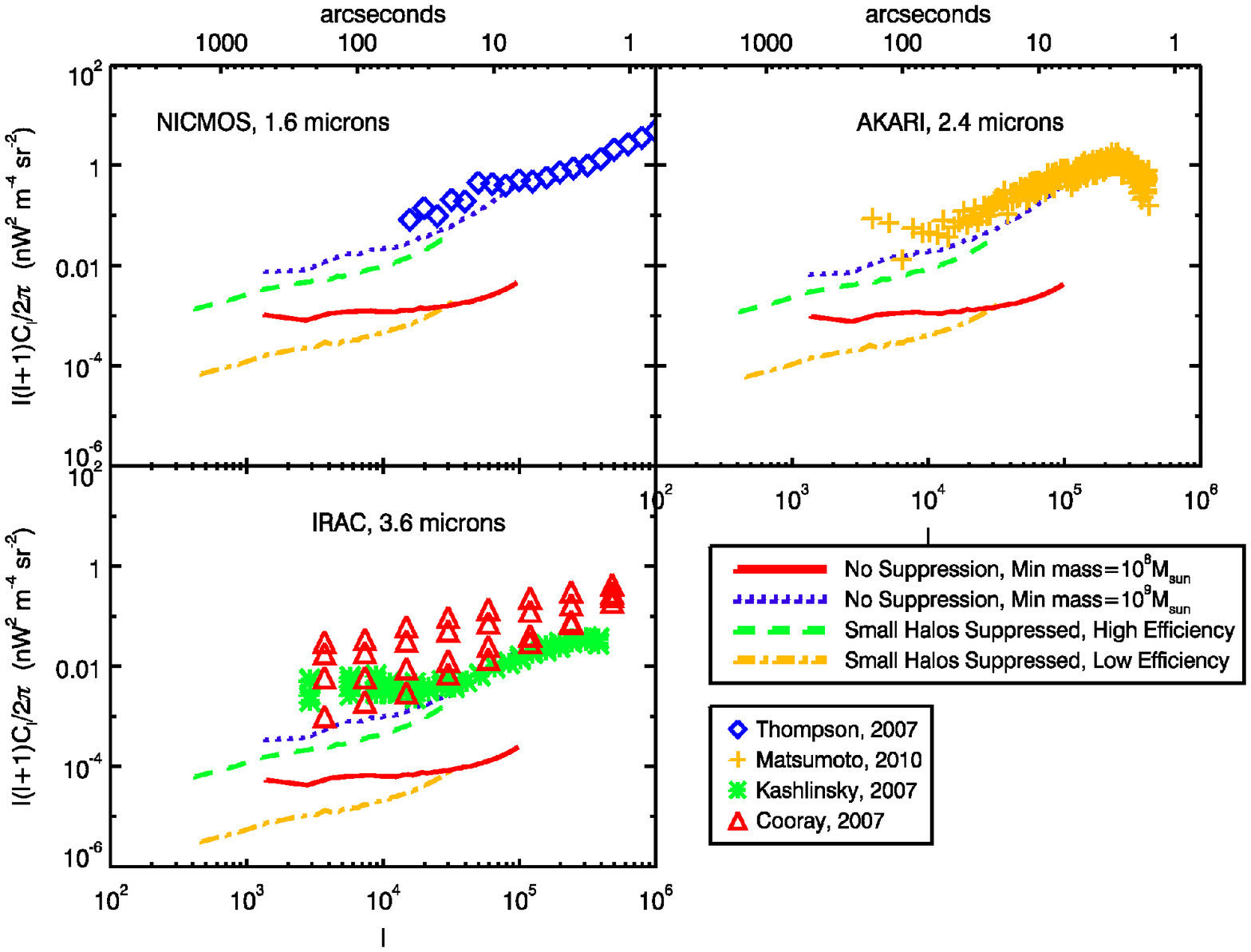}
  \caption{(Left) Angular power spectra of the NIRB fluctuations for sources assumed to be Pop II galaxies with $f_{\mathrm{esc}}=0.1$ and Salpeter IMF (after subtracting the shot-noise contribution), normalized to same amplitude at $l\approx 5000$; (Right) Same model, predictions in various bands compared to observational results. From \cite{fernandez/etal:2012}.  }
\end{figure}

\begin{figure}
  \includegraphics[height=.25\textheight]{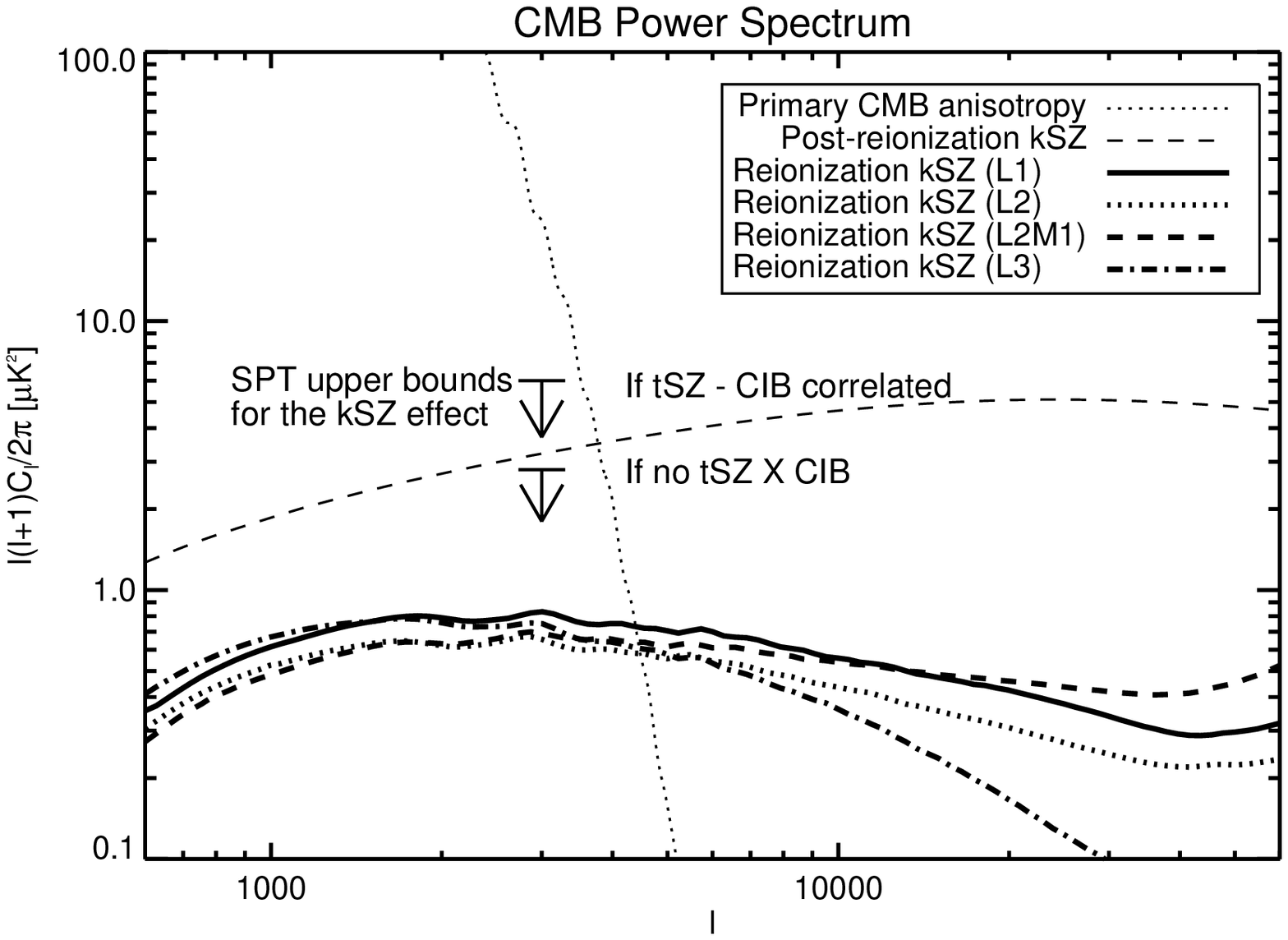}
  \caption{  Angular fluctuation power spectra for the kSZ effect, including contributions from post-reionization and the EOR for several reionization cases, as labelled (see Table 1), along with the SPT upper limits from \cite{2012ApJ...755...70R}.  From Park et al. 2012, in prep.}
\end{figure}

\emph{Can near-IR background fluctuations tell us which galaxies reionized the universe?}
The same stars that supply the ionizing radiation that escapes from galactic sources of reionization emit radiation below the H ionization threshold which escapes directly from the galaxies (in the absence of dust), and the absorbed fraction ($1 - f_{\mathrm{esc}}$) of their ionizing photons is reprocessed into nebular emission that also escapes.  This makes reionization sources also sources of the near-IR background, and their spatial clustering and its evolution contribute angular fluctuations in this background which probes the mass range of halos responsible for reionization, as well.  We have applied our simulations of reionization to predict this fluctuating background in the near-IR, 
previously without the LMACHs and their self-regulation \cite{fernandez/etal:2010},
and now with them \cite{fernandez/etal:2012}.  The talk by Beth Fernandez described this work in more detail.  The angular power spectrum $C_l$ tends to be dominated by the galaxies responsible for \emph{completing} reionization (e.g. $z \sim 6$).  The shape of the angular power spectrum $C_l$ at high $l$ is sensitive to the amount of nonlinear bias of these galaxies relative to the total matter density, and since this bias depends on halo mass, the shape can be used to tell us what galactic halo mass was responsible for completing reionization.   As Figure 5 shows, we find that $C_l$ is steeper for the case with suppression of LMACHs than with LMACHs included \emph{without} suppression.  In all cases, we do not see a \emph{turn-over} toward high $l$ in the shape of $l^2 C_l$.

\emph{Can the kinetic Sunyaev-Zel'dovich effect from patchy reionization tell us which galaxies reionized the universe?}  The kSZ effect is a CMB temperature anistropy induced by electron scattering by free electrons moving along the line-of-sight, and is distinguished from the related thermal SZ effect induced by the scattering by hot electrons in the intracluster gas, by the spectral difference between these two effects.  As an integrated effect over the path of photons from recombination to the present, the kSZ signal is the sum of the contributions from the IGM during the EOR and post-reionization era.  Recent results by the South Pole Telescope (SPT) have detected the tSZ effect and placed an upper limit on the kSZ effect at arcminute scale, $l \sim 3000$ \cite{2012ApJ...755...70R}, which has been interpreted as a constraint on the duration and timing of reionization, imposed by the need to keep the EOR contribution below the upper limits, after subtracting the post-reionization signal expected theoretically, from the upper limit on the total kSZ signal \cite{2012ApJ...756...65Z,2012MNRAS.422.1403M}.  Nondetection by the SPT yields an angular fluctuation power upper limit on the total kSZ signal at $l=3000$ of $D_l < 2.8  \mu\mathrm{K}^2$ ($95\%$ confidence), where $D_l = l(l+1) C_l/ 2 \pi$, but if there is a correlation between galactic emission at these wavelengths and the tSZ effect, this introduces an uncertainty which raises the upper bound to the more conservative level of 6 $\mu\mathrm{K}^2$ \cite{2012ApJ...755...70R}.

Our earlier predictions of the EOR contribution to this kSZ
without LMACHs \cite{iliev/etal:2007b,iliev/etal:2008}, have now been replaced by our new 163 Mpc box simulations of self-regulated reionization, listed in Table 1.
The results are summarized in Table 1 and Figure 6, from Park et al. (2012, in prep).
All cases are easily consistent with the more conservative bound if there is a tSZ-CIB correlation, but they are also all allowed by the tighter bound, too.  Apparently, the global reionization histories for these models include a range of durations which are larger than were found to be allowed by comparison of semi-numerical reionization models with the SPT upper limits by \cite{2012ApJ...756...65Z} and \cite{2012MNRAS.422.1403M}.
Alas, it does not yet seem possible to use the SPT limits on the kSZ fluctuations from patchy reionization to determine which galaxies reionized the universe, even though the results are beginning to place a significant constraint on the history of reionization.

\begin{theacknowledgments}
This study was supported
in part by U.S. NSF grants AST-0708176 and AST-1009799; NASA
grants NNX07AH09G, NNG04G177G, and NNX11AE09G;
and Chandra grant SAO TM8-9009X, and by the Swedish Research Council grant 2009-4088.  K.A. was supported in part by NRF grant funded by
the Korean government MEST (Nos. 2009-0068141, 2009-
0076868, 2012R1A1A1014646, and 2012M4A2026720). I.T.I.
was supported by The Southeast Physics Network (SEPNet)
and the Science and Technology Facilities Council grants
ST/F002858/1 and ST/I000976/1. The authors acknowledge the TeraGrid and the Texas Advanced Computing Center (TACC) at The University of Texas at Austin (URL:
http://www.tacc.utexas.edu), and the Swedish National Infrastructure for Computing (SNIC) resources at HPC2N (UmeŒ,
Sweden) for providing HPC and visualization resources that
have contributed to the results reported within this paper. 
\end{theacknowledgments}

\bibliographystyle{aipproc}   

\bibliography{kyoto}

\end{document}